\documentclass[aps,pra,floatfix,twocolumn,superscriptaddress]{revtex4-1}
\usepackage{amssymb,amsthm,amsmath,amsfonts}
\usepackage{graphicx,ulem,mathptmx}
\usepackage[pdftex,dvipsnames,usenames]{xcolor}
\usepackage[colorlinks=true,urlcolor=blue,citecolor=blue,linkcolor=blue]{hyperref}

\begin{document}

\title{Measuring coherence of quantum measurements}

\author{Valeria Cimini}
\affiliation{Dipartimento di Scienze, Universit\`{a} degli studi Roma Tre, Via della vasca Navale 84, 00146 Rome, Italy}

\author{Ilaria Gianani}
\affiliation{Dipartimento di Scienze, Universit\`{a} degli studi Roma Tre, Via della vasca Navale 84, 00146 Rome, Italy}
\affiliation{Dipartimento di Fisica, Sapienza Universit\`{a} di Roma, Piazzale Aldo Moro 5, 00185 Rome, Italy}

\author{Marco Sbroscia}
\affiliation{Dipartimento di Scienze, Universit\`{a} degli studi Roma Tre, Via della vasca Navale 84, 00146 Rome, Italy}

\author{Jan Sperling}\email{jan.sperling@uni-paderborn.de}
\affiliation{Integrated Quantum Optics Group, Applied Physics, University of Paderborn, 33098 Paderborn, Germany}

\author{Marco Barbieri}
\affiliation{Dipartimento di Scienze, Universit\`{a} degli studi Roma Tre, Via della vasca Navale 84, 00146 Rome, Italy}
\affiliation{Consiglio Nazionale delle Ricerche, Largo E. Fermi 6, 50125 Florence, Italy}

\date{\today}

\begin{abstract}
	The superposition of quantum states lies at the heart of physics and has been recently found to serve as a versatile resource for quantum information protocols, defining the notion of quantum coherence.
	In this contribution, we report on the implementation of its complementary concept, coherence from quantum measurements.
	By devising an accessible criterion which holds true in any classical statistical theory, we demonstrate that noncommutative quantum measurements violate this constraint, rendering it possible to perform an operational assessment of the measurement-based quantum coherence.
	In particular, we verify that polarization measurements of a single photonic qubit, an essential carrier of one unit of quantum information, are already incompatible with classical, i.e., incoherent, models of a measurement apparatus.
	Thus, we realize a method that enables us to quantitatively certify which quantum measurements follow fundamentally different statistical laws than expected from classical theories and, at the same time, quantify their usefulness within the modern framework of resources for quantum information technology.
\end{abstract}

\maketitle

%%%%%%%%%%%%%%%%%%%%%%%%%%%%%%%%%%%%%%%%%%%%%%%%%%%%%%%%%%%%%%%%%%%%%%%%%%%%%%%%%%%%%%%
% Introduction
%%%%%%%%%%%%%%%%%%%%%%%%%%%%%%%%%%%%%%%%%%%%%%%%%%%%%%%%%%%%%%%%%%%%%%%%%%%%%%%%%%%%%%%

\section{Introduction}
	Quantum interference phenomena are a key property that enable us discern classical physics from the quantum realm \cite{BCP14,WY16,TKEP17,BDW17,WTWYKXLG17}.
	Different forms of quantum coherence constitute the basis for a variety of notions of nonclassicality, such as entanglement which is a result of nonlocal superpositions \cite{VS14,SSDBA15,KSP16,CH16,Qetal18}.
	The application of quantum coherence as a resource for quantum information protocols recently gained a lot of attention (see Refs. \cite{ABC16,SAP17,CG18} for introductions) because it connects fundamental question about the physical nature with practical aspects of upcoming quantum technologies.

	In order to show how classical expectations are superseded by quantum physics, a number of measurable criteria have been proposed.
	Most prominently, Bell's inequality \cite{B64} enables us to show that local hidden-variable models do not sufficiently describe general correlations between quantum systems.
	More generally, the concept of contextuality provides a broadly applicable approach which demonstrates the superiority of quantum-mechanical joint probabilities over their classical counterparts \cite{HWVE14,R13}.
	The underlying constraints for both examples provide criteria which were derived in a classical picture to demonstrate how quantum physics overcomes classical limitations; see Ref. \cite{ZXXZSKZ19} for a recent experiment.

	Conversely, the resource-theoretic framework of quantum coherence is already formulated in the quantum formalism and quantifies the operational usefulness of superpositions when compared to quantum-statistical ensembles \cite{ABC16,SAP17,CG18}.
	In fact, quantum superpositions themselves can directly serve as a measure of quantum coherence \cite{A06,A14,SV15}.
	Moreover, the general concept of quantum coherence encompasses previous notions of quantumness, e.g., quantum-optical nonclassicality as formulated by Glauber and others \cite{G63,TG65,M86,MW95,VW06}, and thereby renders it possible to distinguish classical interference phenomena from coherence effects genuine to quantum physics.

	Except for some recent approaches with remarkable implications \cite{YDXLS17,BKB18,SL18,SSC19,TR19,GMSG19}, the state-based approach to quantum coherence does not address the quantumness of measurement itself \cite{WM08}.
	However, there can be no doubt that the understanding of the nature of measurements is vital to the fundamentals of physics and its practical applications, e.g., allowing us to implement measurement-based quantum computation \cite{NMDB06,BBDRN09}.
	Other scenarios in which the coherence of measurements becomes essential relate to the manipulation and preparation of quantum states \cite{HF16,CSRBAL16,MZFL16,G19}, conditional quantum correlations \cite{SBDBJDVW16,ASCBZV17}, and questions concerning the collapse of the wave function \cite{YDXLS17,FTZWF15,KKYMTYSLM16,S18,Xetal19}.
	The other way around, Heisenberg's seminal uncertainty relation \cite{H27} poses a fundamental precision limitation to quantum measurements of multiple observables \cite{SPB17,YBPM17,DHSK19}, which is not the case in classical models.
	Thus, an experimentally accessible distinction between classical and quantum statistics, based on the outcomes of measurements, is vital for many applications.
	While some measurements have been performed, for example, to confirm the noncommutativity of certain observables \cite{PZKB07,ZPKJB09}, a general connection between the quantumness of measurements and the state-based notion of quantum coherence, together with its experimental certification, is still missing.

	In this contribution, we close this gap between the theory of quantum coherence of states and experiments with incompatible quantum measurements.
	To derive our experimentally accessible and generally applicable criteria, we first perform a derivation in a purely classical framework;
	second, we relate our findings to quantum coherence of measurements.
	Then, we apply our technique to data obtained in our experiment of polarization measurements of photons, detecting one qubit of information.
	Our results not only verify with high statistical significance if and when a classical interpretation of a measurement ultimately fails in quantum systems, but it also provides a quantifier of the measurement-based quantum coherence.
	Thus, we provide and implement a practical tool to study the fundamentals and application-oriented properties of quantum measurements.

%%%%%%%%%%%%%%%%%%%%%%%%%%%%%%%%%%%%%%%%%%%%%%%%%%%%%%%%%%%%%%%%%%%%%%%%%%%%%%%%%%%%%%%
% Theory
%%%%%%%%%%%%%%%%%%%%%%%%%%%%%%%%%%%%%%%%%%%%%%%%%%%%%%%%%%%%%%%%%%%%%%%%%%%%%%%%%%%%%%%

\section{Classical law of total probabilities}
	Like the approach by Bell and others, let us formulate our classical constraints solely based on universally valid features of classical statistics.
	For this reason, we consider a probability distribution $P$, where $P(x)$ and $P(y)$ are the probabilities to measure the outcomes $x$ and $y$ for two random variables.
	Further, the probability to measure $y$ after a measurement of $x$ is given by the conditional probability $P(y|x)=P(y,x)/P(x)$, where $P(y,x)$ is the joint probability for the given outcomes.
	Consequently, the probability to detect $y$ regardless of the prior outcome $x$ is given by $P'(y)=\sum_x P(y|x) P(x)$.
	According to the law of total probability \cite{B69,S95}, we have
	\begin{equation}
		\label{eq:TotalProb}
		P(y)\stackrel{\text{cl.}}{=}P'(y)
	\end{equation}
	for any classical system.
	It is worth emphasizing that the law of total probability applies to any classical model even if the measurement is not an ideal one.

	Using the classical identity \eqref{eq:TotalProb}, we can now formulate a variance-based constraint for classical statistics,
	\begin{equation}
		\label{eq:TotalVar}
		\mathbb V_{P(y)}[y]
		\stackrel{\text{cl.}}{=}\mathbb V_{P'(y)}[y],
	\end{equation}
	where $\mathbb V$ denotes the variance.
	This classical relation is known as the law of total variance \cite{B69,S95} and follows from the decomposition $\mathbb V_{P'(y)}[y]=\mathbb E_{P(x)}[\mathbb V_{P(y|x)}[y]]+\mathbb V_{P(x)}[\mathbb E_{P(y|x)}[y]]$, which is based on the construction of $P'$ via conditional probabilities and where $\mathbb E$ denotes the mean value.
	A violation of the classically universal law in Eq. \eqref{eq:TotalVar} certifies the incompatibility of the measurement with classical statistics.
	It is worth emphasizing that beyond the second-order criterion \eqref{eq:TotalVar}, more sophisticated generalizations are possible using Eq. \eqref{eq:TotalProb}.

\section{Relation to quantum coherence}
	Let us now establish the relation of the above criterion to the notion of quantum coherence.
	For this reason, we identify an observable, represented through the operator $\hat x$, to serve as our incoherent gauge when compared to a second, general observable $\hat y$.
	The decomposition of those observables reads $\hat x=\sum_x x\,\hat\Xi_x$ and $\hat y=\sum_y y\,\hat\Pi_y$, using the positive operator-valued measures $\{\hat\Xi_x\}_x$ and $\{\hat\Pi_y\}_y$.

	Measuring the outcome $x$ is achieved with the probability $P(x)=\mathrm{tr}(\hat\rho\hat\Xi_x)=\langle \hat\Xi_x\rangle_{\hat\rho}$ and leaves us with a post-measurement state $\hat\rho_x=\hat\Xi_x^{1/2}\hat\rho\hat\Xi_x^{1/2}/P(x)$.
	In analogy to the classical case, we now ignore the first outcome, resulting in
	\begin{equation}
		\label{eq:Incoherent}
		\hat\rho'=\sum_x P(x)\hat\rho_x=\sum_x \hat\Xi_x^{1/2}\hat\rho\hat\Xi_x^{1/2}.
	\end{equation}
	For our purpose, it is now convenient to define that a state is incoherent if the map in Eq. \eqref{eq:Incoherent} leaves the state unchanged, i.e., $\hat\rho\mapsto\hat\rho'=\hat\rho$.
	Conversely, quantum coherence is given by $\hat\rho'\neq\hat\rho$.
	In this sense, $\hat\rho\mapsto\hat\rho'$ is a so-called strictly incoherent operation \cite{WY16,YMGGV16}.
	In general, assessing coherence demands a choice of a preferred basis on grounds of physical considerations.
	Here, it is motivated through a detection of $\hat x$ because, in itself, it does have a completely classical model in terms of the measured statistics $P(x)$.

	For comparing the two cases, the measurement of $\hat y$ without and with a prior measurement of $\hat x$ yields
	\begin{equation}
		\mathbb V_{P(y)}[y]=\langle (\Delta \hat y)^2\rangle_{\hat\rho}
		\text{ and }
		\mathbb V_{P'(y)}[y]=\langle (\Delta \hat y)^2\rangle_{\hat\rho'},
	\end{equation}
	respectively, corresponding to the variances for the previously discussed classical case.
	Here, however, the classical law of total variances does not apply, and we can find $\langle (\Delta \hat y)^2\rangle_{\hat\rho}\neq\langle (\Delta \hat y)^2\rangle_{\hat\rho'}$ in the presence of quantum coherence, $\hat\rho\neq\hat\rho'$.

	It is worth emphasizing that our approach does indeed measure the incompatibility of the performed measurements because $[\hat\Xi_x,\hat\Pi_y]=0$ ($\forall x,y$) implies $P(y)=P'(y)$, regardless of the coherence of the initial state $\hat\rho$ \cite{CommentProof1}.
	Therefore, when the classical constraint \eqref{eq:TotalVar} is violated, we can directly infer that the quantum measurement $\hat y$ exhibits quantum coherence with respect to the detection of $\hat x$.
	In an ideal scenario, where the measurements are represented through orthonormal bases, i.e., $\hat\Xi_x=|x\rangle\langle x|$ and $\hat\Pi_y=|y\rangle\langle y|$, this means that the measurement of $\hat y$ is not described through incoherent mixtures ($\hat\Pi_y\neq \sum_x q_{y,x}|x\rangle\langle x|$);
	rather, it requires quantum superpositions ($|y\rangle=\sum_x c_{y,x}|x\rangle$), i.e., quantum coherence in the measurement operators.
	More specifically, our intermediate definition of the coherence of the states, based on the measurement of $\hat x$ [cf. Eq. \eqref{eq:Incoherent}], actually serves as a proxy to infer the quantum coherence of the second measurement $\hat y$ when compared to $\hat x$.
	Consequently, we have formulated an observable criterion that assesses the quantum coherence of measurements and which is based off of the classical law of total probabilities.

%%%%%%%%%%%%%%%%%%%%%%%%%%%%%%%%%%%%%%%%%%%%%%%%%%%%%%%%%%%%%%%%%%%%%%%%%%%%%%%%%%%%%%%
% Experiment
%%%%%%%%%%%%%%%%%%%%%%%%%%%%%%%%%%%%%%%%%%%%%%%%%%%%%%%%%%%%%%%%%%%%%%%%%%%%%%%%%%%%%%%

\section{Implementation}
	We explore the previously devised concepts for qubits.
	While our approach applies to arbitrary system, qubits are fundamental quantum objects as they represent the basic unit of quantum information science \cite{KL98,NMDB06,NC00}.
	Our qubits are encoded in the polarization of single photons.
	The preferred basis is given by the horizontal ($H$) and vertical ($V$) polarization, also defining the reference measurement $\hat x=-|H\rangle\langle H|+|V\rangle\langle V|=\left[\begin{smallmatrix}-1&0\\0&1\end{smallmatrix}\right]$.
	The states we prepare take the general form
	\begin{equation}
		\label{eq:State}
		\hat\rho=\begin{bmatrix}
			1-p &
			\sqrt{p(1-p)}\gamma \\
			\sqrt{p(1-p)}\gamma &
			p
		\end{bmatrix},
	\end{equation}
	where $p\in[0,1]$ indicates the population unbalance between the two levels, and the parameter $\gamma\in[0,1]$ determines the coherence in the state when compared to its incoherent counterpart, $\hat\rho'=(1-p)|H\rangle\langle H|+p|V\rangle\langle V|$, cf. Eq. \eqref{eq:Incoherent}.
	Note that $\gamma$ can be additionally equipped with a complex phase factor, $e^{i\phi}$, to account for the most general case of a qubit;
	this, however, does not lead to any conceptional advantage and is, therefore, fixed to one ($e^{i\phi}=1$) in our treatment.
	The qubit is subjected to two consecutive measurements.
	The first one measures the Pauli-$z$ operator (here, denoted as $\hat x$);
	the second one measures an arbitrary observable 
	\begin{equation}
		\label{eq:Measurement}
		\hat y
		=\cos\theta
		\begin{bmatrix}
			-1 & 0 \\
			0 & 1
		\end{bmatrix}
		+\sin\theta\begin{bmatrix}
			0 & 1 \\
			1 & 0
		\end{bmatrix},
	\end{equation}
	parametrized through the angle $\theta$.

	In our experiment, we prepare linear polarization states from $H$-polarized photons by means of a half-wave plate (HWP) at an angle $\alpha$---hence, $p=\sin^2(2\alpha)$.
	The statistics for mixed states is obtained by inputting states corresponding to the setting $+\alpha$ and $-\alpha$ with the respective weights $w_+,w_-\geq0$, chosen in such a way that ${w_++w_-=1}$ and ${w_+-w_-=\gamma}$ hold true.

	In order to implement the $\hat x$ measurement without destroying the signal photon, we couple the photon with an ancillary meter by means of a controlled sign gate \cite{L05,KSWUW05,OHTS05}.
	This is constituted by a beam splitter with polarization-dependent transmittivities, $T_H=1$ and $T_V=1/3$.
	Two-photon nonclassical interference occurs selectively on the beam splitter only for the vertical components of the signal and the meter photon polarizations.
	These consequently acquire a $\pi$ phase shift with respect to the other three terms when post-selecting those events in which the two photons emerge on distinct outputs of the beam splitter---this then demands to register only coincidence events between the arms. 
	In our setup, we encounter slight imperfections, such as $T_H = 98.5\%\neq1$ and $T_V = 32.4\%\neq1/3$.
	In order to reduce the impact of leakage of the horizontal polarization, the polarization modes of the meter photon are kept spatially separated by embedding the controlled gate in a Sagnac interferometer.
	Two other identical beam splitters, rotated by $90^\circ$, are used to balance polarization-dependent loss induced by the first one \cite{RGMSSGB18}.
	Remaining discrepancies between our data and the theoretical modeling can be attributed to reduced visibility of the two-photon interference, as well as the Sagnac interferometer.
	This, however, does not affect the working of our criterion \eqref{eq:TotalVar} as it applies to arbitrary measurements, including their imperfections.

	The action of the described gate can be switched on and off by controlling the polarization of the meter \cite{PBWBR04,MSRGSMPB18,MSRGCPB18}.
	Consider a pure signal state state ($\gamma=1$) arriving at the gate.
	If a $H$-polarized meter is injected, no coupling can occur;
	hence, the joint state remains separable, $\left(\sqrt{1-p}| H \rangle_s+\sqrt{p}| V \rangle_s\right)\otimes|H \rangle_m$.
	No information can be inferred from the meter about the signal.
	If the meter is, however, injected in a diagonal polarization, $|+\rangle_m=\left(|H \rangle_m+|V \rangle_m\right)/\sqrt{2}$, the output two-photon polarization state is entangled, $\sqrt{1-p}| H \rangle_s\otimes | + \rangle_m+\sqrt{p}| V \rangle_s\otimes| - \rangle_m$, due to the phase shift imparted on the gate.
	By measuring the meter in the diagonal basis, one can extract the full information about the $\hat x$ measurement of the original signal state.
	The observable $\hat y$ is then measured conventionally by a HWP at $\beta = \theta/4$ and a polarization beam splitter.

	The experiment is carried out in two steps:
	First, we measure the unperturbed variance, predicted to be $\langle(\Delta\hat y)^2\rangle_{\hat\rho}=1-[(2p-1)\cos\theta+2\sqrt{p(1-p)}\gamma\sin\theta]^2$;
	second, we measure the variance resulting from a prior measurement of $\hat x$, expected to follow $\langle(\Delta\hat y)^2\rangle_{\hat\rho'}=1-(1-2p)^2\cos^2\theta$.
	In order to account for experimental artifacts, both measurements are performed with the photons passing through the gate, with the polarization of the meter set accordingly, and registering coincidences counts.
	In both cases, the polarization of the meter is not analyzed since we are ignoring the outcome $x$, as expressed in Eq. \eqref{eq:Incoherent}.

\section{Results}
	To assess the amount of measurement-induced quantum coherence when compared to the classical constraint \eqref{eq:TotalVar}, it is convenient to consider the following difference:
	\begin{equation}
		\label{eq:Difference}
		\Delta V=\mathbb V_{P'(y)}[y]-\mathbb V_{P(y)}[y]=\langle(\Delta\hat y)^2\rangle_{\hat\rho'}-\langle(\Delta\hat y)^2\rangle_{\hat\rho}.
	\end{equation}
	Because of Eq. \eqref{eq:TotalVar}, a significant deviation from $\Delta V=0$ is our figure of merit to quantify the amount of coherence.

\begin{figure}
	\includegraphics[width=\columnwidth]{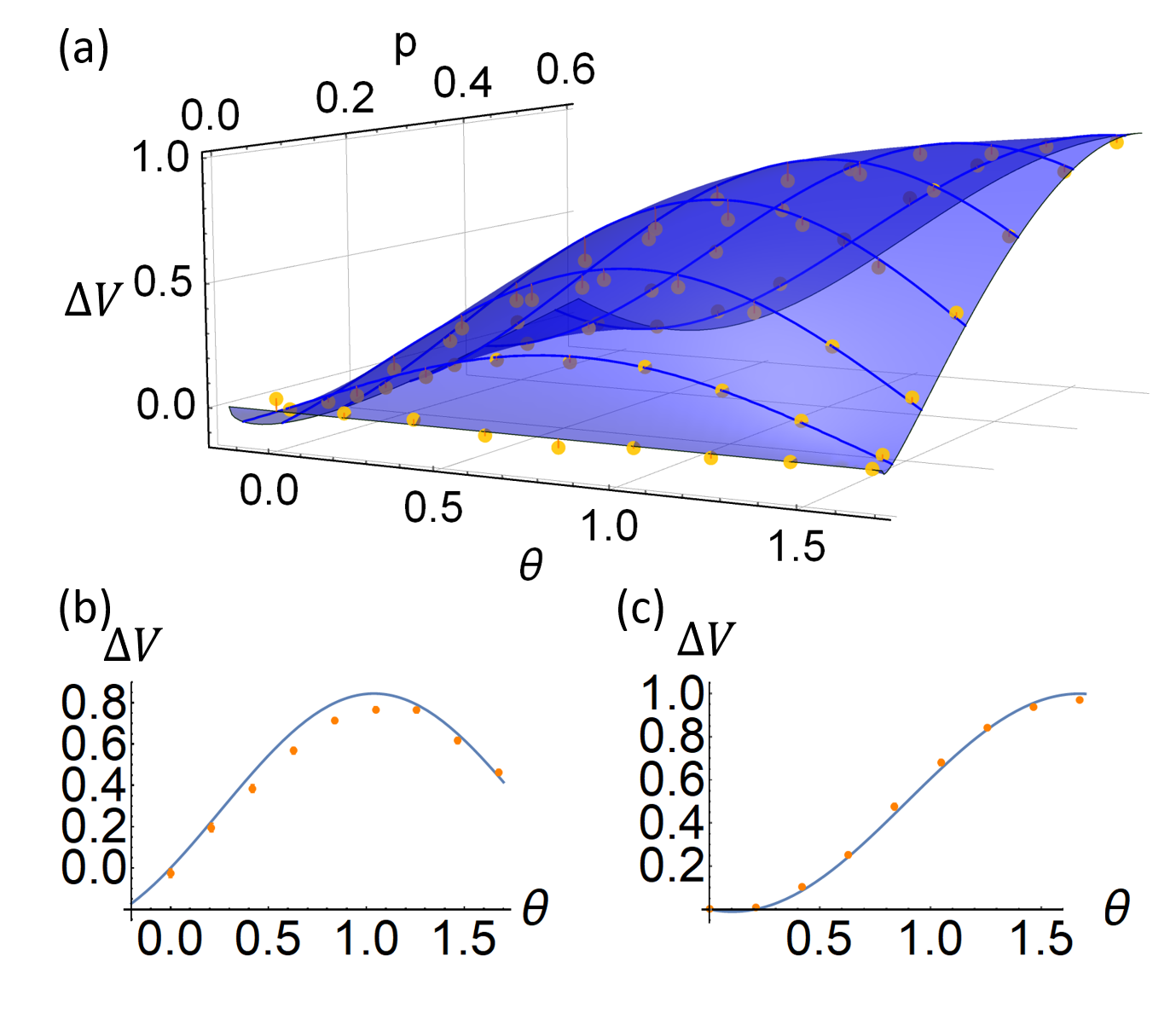}
	\caption{
		Variance difference, representing the deviation from our classical constraint, $\Delta V=0$ [Eq. \eqref{eq:Difference}], between the two measurement configurations.
		(a) The surface depicts the expected theoretical behavior when varying $p$ and $\theta$;
		the points show the experimental data.
		(b) Cut of the plot (a) for $p= 0.165$.
		(c) Cut of the plot (a) graph for $p= 0.552$.
		Error bars, typically smaller than the bullet points, are derived from Poissonian statistics of the count rates;
		the same applies to all following plots.
	}\label{fig:pure}
\end{figure}

	Figure \ref{fig:pure}(a) shows the measured deviation of $\Delta V$ as a function of $\theta$ and $p$ for a measurement of $\hat y$ [Eq. \eqref{eq:Measurement}] for pure states [$\gamma = 1$ in Eq. \eqref{eq:State}].
	As one can observe, our data are in good agreement with the quantum-mechanical model of the detection processes.
	We also confirm that for $\theta\approx0$, i.e., $[\hat x,\hat y]=0$ [Eq. \eqref{eq:Measurement}], no coherence can be observed ($\Delta V=0$) regardless of the input state as predicted in the theory part.
	Moreover, we can rule out that our system can be mimicked by any classical model of a measurement as $\Delta V$ significantly deviates from zero in almost all other cases.
	For specific values of $p$, the deviation from the classical bound is shown in Figs. \ref{fig:pure}(b) and \ref{fig:pure}(c).
	In particular, Fig. \ref{fig:pure}(c) certifies that the maximal violation is obtained for $\theta\approx\pi/2$, which corresponds to a detection of $\hat y$ that is a Pauli-$x$ measurement, and thus maximally incompatible with the reference measurement $\hat x$, the Pauli-$z$ operator.

\begin{figure}
	\includegraphics[width=\columnwidth]{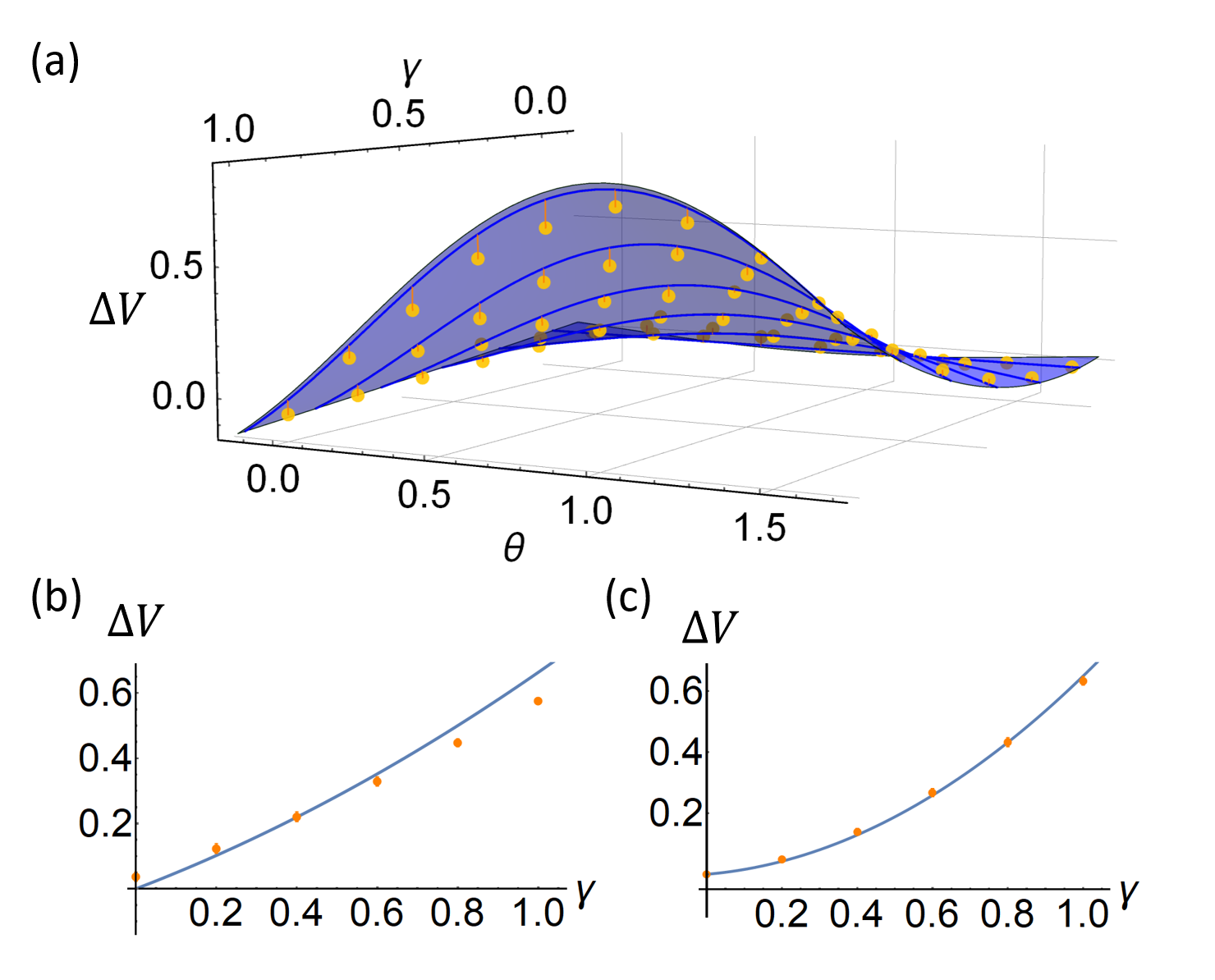}
	\caption{
		Variance difference between the two measurement configurations [Eq. \eqref{eq:Difference}].
		(a) The surface is the expected theoretical behavior as a function of $\gamma$ and $\theta$;
		the points depict our data.
		(b) Cut of the plot (a) along $\theta= 36^\circ$.
		(c) Cut of the plot (a) along $\theta=84^\circ$.
	}\label{fig:mixed}
\end{figure}

	In addition, we explore mixed states to probe the quantum coherence between the measurements in Fig. \ref{fig:mixed}(a).
	For this purpose, we fix the value of $p=\sin^2(2\alpha)$ at $\alpha = 12^{\circ}$ and study the difference of the variances as a function of $\gamma$ [Eq. \eqref{eq:State}].
	We can observe that, in general, the highest purity ($\gamma\to1$) yields the most significant verification of quantum coherence, $\Delta V\neq0$, which also represents the scenario with the highest coherence of the probe state $\hat\rho$.
	Again, Figs. \ref{fig:mixed}(b) and \ref{fig:mixed}(c) show the cuts with the optimal deviations $\Delta V$ from the classical bound zero.

\begin{figure}
	\includegraphics[width=\columnwidth]{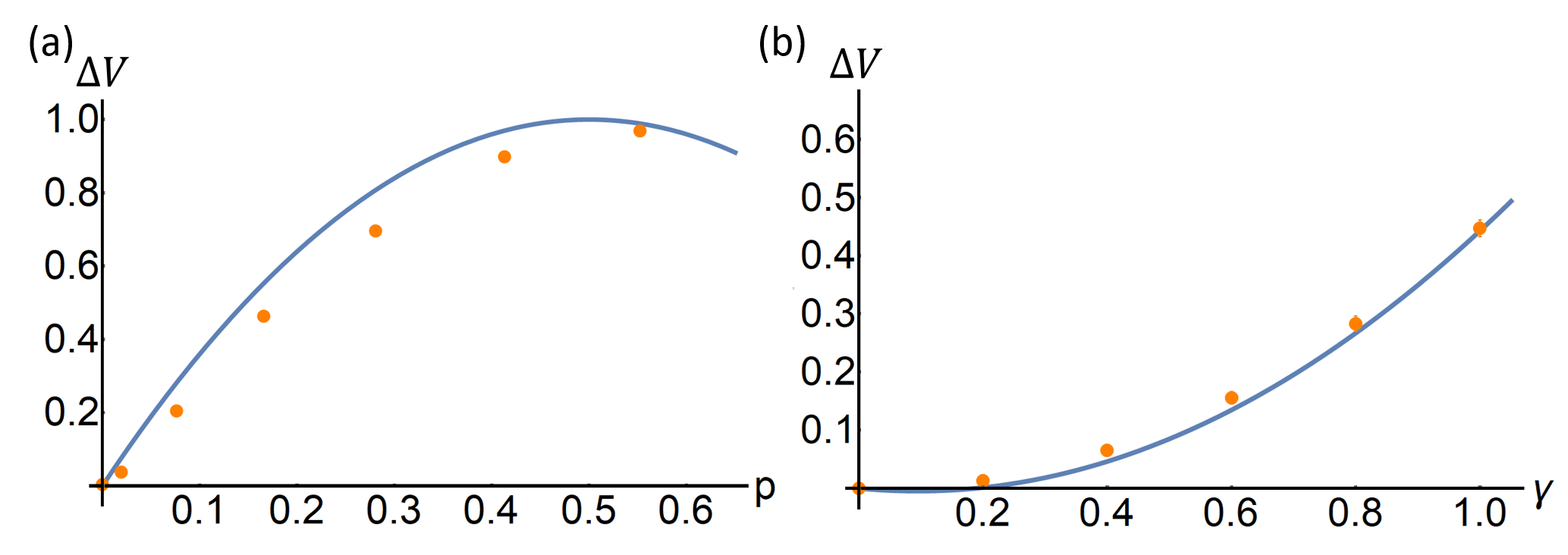}
	\caption{
		Maximal violation of $\Delta V=0$ for $\theta\approx\pi /2$ as a function of $p$ and $\gamma$ in plots (a) and (b), respectively.
		Plot (a) is expected to be symmetric with respect to $p=1/2$ in the ideal case.
	}\label{fig:max}
\end{figure}

	Finally, we can also measure the maximal deviation of the state $\hat\rho$ prior to the measurement $\hat x$ and $\hat\rho'$ after the detection took place \cite{SW18,VMQ19}, cf. Eq. \eqref{eq:Incoherent}.
	The result is shown in Fig. \ref{fig:max} as functions of $p$ and $\gamma$, defining the prepared state in Eq. \eqref{eq:State}.
	The shown results enable us to quantify the measurement-induced decoherence because one can straightforwardly prove \cite{CommentProof2} that the square of trace distance between $\hat\rho$ and $\hat\rho'$ is identical to $\Delta V$ for $\theta\approx\pi/2$.
	For instance, we can conclude from Figs. \ref{fig:max}(a) and \ref{fig:max}(b) that the maximal decoherence occurs for $p\approx 1/2$ and $\gamma\approx 1$.
	This corresponds to a maximally coherent input state, $\hat\rho=|\psi\rangle\langle\psi|$ with $|\psi\rangle=(|H\rangle+|V\rangle)/\sqrt 2$, that is converted into a maximally incoherent one, $\hat\rho'=(|H\rangle\langle H|+|V\rangle\langle V|)/2$, through the detection of $\hat x$.

%%%%%%%%%%%%%%%%%%%%%%%%%%%%%%%%%%%%%%%%%%%%%%%%%%%%%%%%%%%%%%%%%%%%%%%%%%%%%%%%%%%%%%%
% Conclusions
%%%%%%%%%%%%%%%%%%%%%%%%%%%%%%%%%%%%%%%%%%%%%%%%%%%%%%%%%%%%%%%%%%%%%%%%%%%%%%%%%%%%%%%

\section{Discussion}
	In summary, we formulated and implemented a method that enables us to certify quantum coherence between two measurements.
	We applied the law of total probabilities (and variances) to formulate conditions that apply to all classical measurements.
	The translation to the quantum domain enabled us to violate these classical requirements, and thereby we revealed a connection to the notion of quantum coherence between measurements.
	We confirmed our theory by probing the quantumness of different and essential qubit measurements, encoded in the polarization of photons.
	This allowed us to experimentally verify the fundamental incompatibility of quantum measurements with classical statistical models on a quantitative basis.
	Furthermore, we were able to assess the measurement-induced decoherence which occurs when a measurement is performed on a quantum system.

	Our studies reveal fundamental and application-oriented properties genuine to the quantum description of measurements.
	First, we confirmed---with an easily accessible, alternative approach and high statistical significance---that the quantum statistics of measurements has fundamentally different properties than expected from any classical perspective.
	Second, we were able to connect the resource-theoretic notion of quantum coherence of quantum states to the coherence between two measurement scenarios.
	Specifically, one measurement defines a classical reference, the incompatibility of this reference with the employed state and the second measurement then lead to quantum effects beyond classical physics.
	In this scenario, the coherence of the state serves as a medium to prove that the description of the second measurement requires quantum superpositions since for commuting observables, any quantum coherence of the state become meaningless.
	This further demonstrates that, in quantum physics, it makes a profound difference if one measures a second observable in the context of preceding one or not---even if one is ignorant to the outcome of the first detection event.
	Note that the role of the first and second measurements is fixed by reasons of experimental practicality and can be exchanged in our treatment without affecting any of our general observations.

	Our approach also enables us to quantify the loss of coherence as a result of the alteration of a state after a quantum-measurement process took place, relating to the collapse of the wave function.
	In particular, we show that intervening with a measurement has a disruptive action on the quantum information carried by the state's coherence.
	Indeed, a prior measurement cancels the presence of coherence in the state, affecting a subsequent measurement, which is also the basis of our quantumness criteria.
	Furthermore, our second-order criterion can be straightforwardly extended to higher order correlations and other nonlinear statistical quantifiers, such as the entropy, outlining possible generalizations.
	Moreover, our method can be also generalized to compare more than two measurements in a pairwise manner or when measured successively.
	We also want to stress that our approach applies to discrete and continuous variables alike, and its purely classical derivation does not presuppose any knowledge about quantum physics, such as notions of eigenstates, measurement operators, collapse of the wave function, etc.
	This is in contrast to other attempts to quantify the coherence of measurements that require a quantum description.

	It is also worth emphasizing that measurement-based quantum protocols rely on realizing measurements which are incompatible.
	Here, we were able to assess the quantum coherence between such measurements to quantify this resource of incompatibility, which is analogous to the requirements on quantum protocols which exploit the coherence of the state.
	Based on our method interpreted as a means to quantify the measurement-induced decoherence, we have additionally demonstrated how to infer the coherence in the trace distance of a qubit state via the performed measurements, providing the necessary information for the success of certain quantum tasks \cite{TEZP19,OB19}.
	Furthermore, the variance-based form of our criterion enables us to predict the precision to estimate a quantity in settings which consists of noncommuting measurements, which is useful, e.g., when comparing classical and quantum metrology \cite{MT19}.
	Thus, we also provide a useful tool to quantify the coherence of measurements for practical purposes.

%%%%%%%%%%%%%%%%%%%%%%%%%%%%%%%%%%%%%%%%%%%%%%%%%%%%%%%%%%%%%%%%%%%%%%%%%%%%%%%%%%%%%%%
% Akn & References
%%%%%%%%%%%%%%%%%%%%%%%%%%%%%%%%%%%%%%%%%%%%%%%%%%%%%%%%%%%%%%%%%%%%%%%%%%%%%%%%%%%%%%%

\begin{acknowledgments}
	M. S. acknowledges support from the ADAMO project of Distretto Tecnologico Beni e Attivit\`a Culturali, Regione Lazio.
	The Integrated Quantum Optics group acknowledges financial support from the Gott\-fried Wilhelm Leibniz-Preis (Grant No. SI1115/3-1).
\end{acknowledgments}


\begin{thebibliography}{99}
\bibitem{CommentProof2}
	The trace distance between $\hat\rho$ [Eq. \eqref{eq:State}] and $\hat\rho'$ [Eq. \eqref{eq:State} for $\gamma=0$] reads $\|\hat\rho-\hat\rho'\|=2\gamma\sqrt{p(1-p)}$.
	For $\theta=\pi/2$ in Eq. \eqref{eq:Measurement}, we directly get $\Delta V=4\gamma^2p(1-p)$.
	A more general relation, including complex $\gamma$, can be found in Ref. \cite{S18}.
\end{thebibliography}
\end{document}